\documentstyle[aps,prl,amssymb,amsmath,graphicx]{revtex}

\begin{document}
\draft
\twocolumn
\narrowtext
\wideabs{
  \title{Channel Estimation with Noisy Entanglement}
  \author{Thorsten C.\ Bschorr, Dietmar G.\ Fischer, Matthias Freyberger} 
  \address{Abteilung f\"ur Quantenphysik, Universit\"at Ulm, D-89069 Ulm, Germany} 
  \maketitle
  \begin{abstract}
  We analyze the Pauli--channel estimation with mixed nonseparable
  states. It turns out that within a specific range entanglement can serve as a
  nonclassical resource. However, this range is rather small, that is entanglement is not very
  robust for this application.
  We further show that Werner states yield the best result of all Bell diagonal states with the same
  amount of entanglement.
  \end{abstract}
}

\newcommand{\psimi}{|\psi^-\rangle}
\newcommand{\psipl}{|\psi^+\rangle}
\newcommand{\phimi}{|\phi^-\rangle}
\newcommand{\phipl}{|\phi^+\rangle}
\newcommand{\bpsimi}{|\psi^-\rangle\langle\psi^-|}
\newcommand{\bpsipl}{|\psi^+\rangle\langle\psi^+|}
\newcommand{\bphimi}{|\phi^-\rangle\langle\phi^-|}
\newcommand{\bphipl}{|\phi^+\rangle\langle\phi^+|}
\newcommand{\psimn}{|\psi_{m,n}\rangle}
\newcommand{\pmn}{p_{m,n}}


\section{Introduction}

Entanglement is the central concept of quantum information processing \cite{qip}. It has an
intriguingly wide range of consequences, starting from fundamental research \cite{fundres} it nowadays
arrives at amazing possible applications like teleportation \cite{teleport}, quantum cryptography
\cite{qucrypt} and quantum computing \cite{qip}. In all these examples entanglement serves as a constituent having no classical
counterpart. It is therefore tempting to treat it as a new kind of resource unknown to classical
physics.

In order to do so we would like to quantify the amount of entanglement needed for a certain
task, in particular, if such a task cannot be carried through with classical means. But besides this
undoubted significance of entanglement no unique measure of it exists. Abstract as well as
operational approaches have been formulated \cite{horodecki,horodecki-rev}. In the present work we do not resolve this important
problem. But we analyze another task, namely the characterization of a quantum channel, which can be
speeded up with the right amount of entanglement.

For a  maximally entangled state it was shown already in \cite{fischer} that it
enhances the fidelity for estimating the parameters of a
Pauli channel when compared to a scheme based on separable quantum states.
It, however, remained unclear how much entanglement is needed for such an enhancement.
Here we extend this discussion to noisy transmissions, i.\,e.\ to mixed nonseparable states. This allows
us to derive a specific ``strength'' of entanglement which is minimally needed to consider it
as a nonclassical resource for this kind of problem. We regard this as another way of operationally quantifying entanglement.
In particular, we show the special role played by the class of Werner states.

Let us first shortly review the basic problem of channel estimation.
The Pauli channel is defined by the action of a superoperator $C$ on the density operator $\hat
\rho$ via
\begin{equation}\label{superoper}
  C(\hat \rho)=\sum \limits_{i=1}^4 p_i \hat\sigma_i \hat\rho \hat\sigma_i^\dagger\;,
\end{equation}
where the Pauli operators $\hat \sigma_i$ classify the different types of errors, namely no error
($\hat \sigma_4=\hat 1$), bit--flip error ($\hat \sigma_1=|0\rangle\langle 1|+|1\rangle\langle 0|$), phase--flip error
($\hat \sigma_3=|0\rangle\langle 0|-|1\rangle\langle 1|$) and the combination of bit-- and phase--flip error
($\hat \sigma_2=i(|1\rangle\langle 0|-|0\rangle\langle 1|)$).
The different errors $\hat \sigma_i$ appear with probabilities
$\vec{p}=(p_1,p_2,p_3)^{\mathrm T}$, whereas with probability $p_4=1-p_1-p_2-p_3$ no error occurs.

In many applications of quantum information processing we have to be aware of $\vec p$, especially,
if we want to correct for the errors that might have occured during a transmission through the
channel. It is therefore important to ask how we can learn something about the probability vector $\vec p$ of
an unknown Pauli channel. We further assume that we use only a finite number of quantum systems to
unravel $\vec p$. Moreover, we would like to know if we can use nonclassical tools, like entanglement,
provided by the quantum domain. Indeed, it was shown \cite{fischer} that the two parties
(Alice and Bob) connected by the noisy channel can learn more efficiently about the channel
parameters $\vec p$, if they estimate them with the help of maximally entangled Bell states.
We shall now extend this result to mixed nonseparable states in order to study the degree of
entanglement needed for an efficient estimation.

This paper is organized as follows. We begin in Sec.\ \ref{sec:EstSchem} with a discussion of the principle
estimation schemes. Then we proceed by comparing the different estimation schemes in Sec.\ \ref{sec:Compar}.
We examine the special role played by Werner states in our estimation
scheme in Sec. \ref{sec:Bell}. Finally, section \ref{sec:Concl} concludes the paper.


\section{Estimation schemes}\label{sec:EstSchem}

We assume that we have a total resource of $R$ qubits to estimate the probability vector $\vec p$ and
that we are
able to prepare $N=R/2$  entangled qubit pairs (ebits) in a Werner state \cite{werner}
\begin{eqnarray}\label{werner}
  \hat\rho_F &=& F \,\bpsimi \nonumber \\ &+& \frac{1-F}{3} \left( \bpsipl + \bphimi + \bphipl \right)\;,
\end{eqnarray}
which is completely characterized by the fidelity $F=\langle\psi^-|\hat\rho_F|\psi^-\rangle$ and the Bell states
$|\psi^\pm\rangle=\frac{1}{\sqrt{2}}(|0\rangle|1\rangle \pm |1\rangle|0\rangle)$ and
$|\phi^\pm\rangle=\frac{1}{\sqrt{2}}(|0\rangle|0\rangle \pm |1\rangle|1\rangle)$.

A Werner state can---roughly speaking---be considered as a mixture of maximally entangled states
due to imperfections (noise) in the preparation or transfer step \footnote{By randomly applying
bilateral rotations it is possible to obtain a Werner state from any general Bell diagonal
state \cite{bennett}.}.
Let us shortly review its basic characteristics.
In the case $F=\frac{1}{4}$, we obtain a totally mixed state $\hat\rho_{F=1/4}=\frac{1}{4}\hat 1$,
which certainly does not yield any information about the channel.
For $F>\frac{1}{2}$ the Werner state has a nonzero negativity of its partial transpose \cite{vidal}
and is therefore nonseparable that is, for $F>\frac{1}{2}$ there is a chance that a Werner state improves the
parameter estimation compared to the separable case. For $F>\frac{1}{8}(3\sqrt{2}+2)\approx 0.78$
the Werner state violates the Bell--CHSH inequality \cite{chsh,peres}
and for $F=1$ we obtain the maximally entangled Bell state $\psimi$, which was already examined in
\cite{fischer}. In the following we therefore restrict ourselves to the domain $\frac{1}{2}< F\le 1$
where $\hat \rho_F$ is nonseparable.

\begin{figure}[htbp]
  \begin{center}
    \includegraphics[width=\columnwidth]{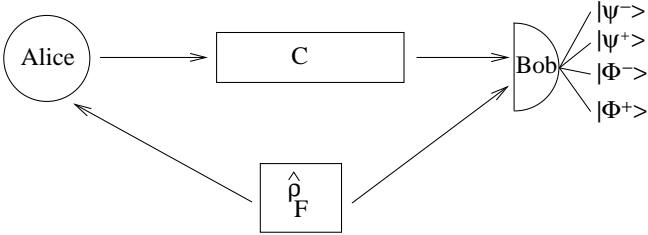}
    \vspace{0.5cm}
    \caption{\label{fig-scheme}
      Principle layout of the estimation scheme. Alice and Bob receive one qubit of an entangled
      state (ebit)
      $\hat\rho_F$. Alice sends her qubit through the channel defined by the superoperator $C$,
      Eq.~\eqref{superoper}, to
      Bob who performs a joint Bell measurement on the two qubits. From the measurement 
      results Bob deduces the parameters of $C$.
    }
  \end{center}
\end{figure}

Let us now turn to the scheme (Fig.~\ref{fig-scheme}) that we are using to determine the channel properties.
First, the ebit has been distributed between Alice and Bob. It is this preparational step which
might change an initially maximally entangled state into the mixture $\hat\rho_F$.
One qubit of each ebit (say Alice's) is sent through the channel while the other (Bob's) is left untouched.
After passing Alice's qubit through the channel we perform a Bell measurement on the output state
\begin{eqnarray}\label{CrhoF}
  C(\hat\rho_F) =
          \left( F-\textstyle\frac{4F-1}{3}(p_1+p_2+p_3)\right)& \bpsimi& \nonumber \\
  + \left( \textstyle\frac{1-F}{3}+\frac{4F-1}{3} p_3 \right)& \bpsipl &\nonumber \\
  + \left( \textstyle\frac{1-F}{3}+\frac{4F-1}{3} p_1 \right)& \bphimi& \nonumber \\
  + \left( \textstyle\frac{1-F}{3}+\frac{4F-1}{3} p_2 \right)& \bphipl&
\end{eqnarray}
and obtain after $N$ measurements the frequencies
\begin{eqnarray}\label{Pest}
  P_{\phimi}^{\mathrm{\,est}} =\frac{i_1}{N} &\;,\;&
  P_{\phipl}^{\mathrm{\,est}} =\frac{i_2}{N} \;,\nonumber \\
  P_{\psipl}^{\mathrm{\,est}} =\frac{i_3}{N} &\;,\;&
  P_{\psimi}^{\mathrm{\,est}} =\frac{i_4}{N}
\end{eqnarray}
where $i_1$ denotes the number of $\phimi$ results, $i_2$ the number of $\phipl$ results, etc. As we
assume that no
qubits are lost during their channel passage, we further have $\sum_{j=1}^4 i_j=N$.

The probabilities of measuring the different Bell states are now connected to the channel parameters
$\vec p$ via
\begin{eqnarray}\label{Ptheo}
P_{\phimi} &=& \textstyle \frac{1-F}{3}+\frac{4F-1}{3}p_1 \nonumber \\
P_{\phipl} &=& \textstyle \frac{1-F}{3}+\frac{4F-1}{3}p_2 \nonumber \\
P_{\psipl} &=& \textstyle \frac{1-F}{3}+\frac{4F-1}{3}p_3 \nonumber \\
P_{\psimi} &=& \textstyle F-\frac{4F-1}{3}(p_1+p_2+p_3)
\end{eqnarray}
as can be seen from Eq.~\eqref{CrhoF}.

Combining Eqs.~\eqref{Pest} and \eqref{Ptheo} enables us to estimate the channel parameters
\begin{equation}
  p_j^{\mathrm{est}} = \frac{3\frac{i_j}{N}+F-1}{4F-1}\:.
\end{equation}
Note that $p_j^{\mathrm{est}}$ can have unphysical negative values which are due to an
imperfect estimation scheme (for $F=1$ we do not run into troubles). In a real experiment one would
treat this as a probability equal zero. However, below we are only interested in the average error
of the estimation $p_j^{\mathrm{est}}$ and hence these negative values do not come into play.
If our estimation scheme is good, this parameters should be close to the actual parameters $\vec p$
of the quantum channel. To quantify this notion of ``closeness'' we introduce the variance of actual
and estimated parameters, $\sum_{j=1}^3(p_j-p_{j}^{\mathrm est})^2$,
to describe the estimation quality. However, this sum only serves to quantify the ``closeness'' of
one single run but we are interested in the average error of our estimation scheme. Therefore we use
the mean quadratic deviation 
\begin{eqnarray}
 \bar g(N,\vec{p}\,)\label{g_general}
         &=& \sum_{i_1+i_2+i_3+i_4=N}\frac{N!}{i_1!i_2!i_3!i_4!} \nonumber \\
         &\times& {P_{\phimi}}^{i_1}\cdot {P_{\phipl}}^{i_2}\cdot {P_{\psipl}}^{i_3}\cdot {P_{\psimi}}^{i_4} \nonumber \\
         &\times& \sum_{j=1}^3(p_j-p_{j}^{\mathrm est})^2
\end{eqnarray}
to quantify the quality of our estimation.
For a Werner state, this average error then becomes
\begin{eqnarray}\label{gWp}
 \bar g_F(N,\vec{p}\,)
         &=& \frac{1}{N}\left(\frac{3}{4F-1}\right)^2 \nonumber \\
         &\times& \sum_{i=1}^3 \Bigl\{ p_i(1-p_i) + \textstyle\frac{4}{3}(1-F)(p_i-\frac{1}{4})(2p_i-1) \nonumber \\
         && \qquad\quad +\textstyle\frac{16}{9}(1-F)^2(p_i-\frac{1}{4})^2 \Bigr\}
\end{eqnarray}
or, if we reexpress everything in the measurement probabilities Eq.~\eqref{Ptheo}, which depend on the channel
probabilities $\vec p$, we obtain
\begin{eqnarray}\label{gWB}
 \bar g_F(N,\vec{p}\,) 
         &=& \left(\frac{3}{4F-1}\right)^2 \cdot\frac{1}{N} \big[ P_{\phimi}(1-P_{\phimi}) \nonumber \\ 
         &+& P_{\phipl}(1-P_{\phipl}) + P_{\psipl}(1-P_{\psipl}) \big]\;.
\end{eqnarray}
For $F=1$ we obtain the result
\begin{eqnarray}\label{g}
 \bar g_{F=1}(N,\vec{p}\,) &=& \frac{1}{N}\left[P_{\phimi}(1-P_{\phimi})+P_{\phipl}(1-P_{\phipl})\right .\nonumber \\
                    &&\quad \left. + P_{\psipl}(1-P_{\psipl})\right]\nonumber \\
                    &=& \frac{1}{N}\left[p_1(1-p_1)+p_2(1-p_2)+p_3(1-p_3)\right]\;, \nonumber \\
\end{eqnarray}
which was already derived in \cite{fischer}.
Moreover the above results Eqs.~\eqref{gWp}--\eqref{g} can be nicely generalized to $d$
dimensions. We shortly present the main steps in Appendix \ref{sec:d-dim-werner}.

In order to compare the estimation error, Eq.~\eqref{gWp}, using nonseparable states to the
separable case, we also shortly review the estimation scheme for separable states.
To determine the error probabilities $\vec p$ Alice prepares uncorrelated qubits in three well
defined reference states,
and sends them independently through the channel to Bob. He finally performs one measurement on each of these qubits.
Note that for this comparison we can restrict ourselves to separable qubits described by pure
separable states, since mixed states will definitely lead to additional noise in the estimation scheme.

For the three different error operators (a) $\hat\sigma_1$, (b) $\hat\sigma_2$ and
(c) $\hat\sigma_3$ of the Pauli channel, Alice prepares the pure states (a) $\frac{1}{2}(\hat 1+\hat\sigma_1)$,
(b) $\frac{1}{2}(\hat 1+\hat\sigma_2)$ and (c) $\frac{1}{2}(\hat 1+\hat\sigma_3)$ respectively and sends them
through the channel. In order to obtain a fair comparison, Alice again only uses a total number of
$R$ qubits and therefore $M=R/3$ qubits for each of the three input states.
Bob measures the operators (a) $\hat \sigma_1$, (b) $\hat \sigma_2$ and (c) $\hat \sigma_3$ 
and uses the corresponding expectation values $\langle \hat\sigma_i \rangle$ to calculate the
parameter vector $\vec p$.
The quality of the estimation which again can be measured using the averaged quadratic deviation
then reads \cite{fischer}
\begin{eqnarray}\label{f}
 \bar{f}(M,\vec{p}\,)
   &=& \frac{3}{2M} \left[p_1(1-p_1)+p_2(1-p_2)+p_3(1-p_3)\right. \nonumber \\
   && \quad\quad - \left. p_1 p_2-p_2 p_3-p_1 p_3 \right]\;.
\end{eqnarray}
In what follows the quantity $\bar{f}$ serves as a reference. In the next section we will show under
which conditions we can improve this error bound by using nonseparable qubits.


\section{Comparison of the different estimation schemes}\label{sec:Compar}

In this section we compare the three different estimation schemes and the corresponding errors, namely
$\bar{f}(M,\vec{p}\,)$, Eq.~\eqref{f}, for separable states,
$\bar{g}_F(N,\vec{p}\,)$, Eq.~\eqref{gWp}, for nonseparable Werner states and
$\bar{g}_{F=1}(N,\vec{p}\,)$, Eq.~\eqref{g}, for maximally entangled states.

In \cite{fischer} it was shown that $\bar{f}(M=R/3,\vec{p}\,) \ge \bar{g}_{F=1}(N=R/2,\vec{p}\,)$ for all
possible parameters $\vec p$. This means that an estimation with prior maximal entanglement is always
superior to an estimation with
separable states. But to what extent does this still hold if we only have our imperfectly entangled
Werner states, Eq.~\eqref{werner}, for estimation? Or, in other words, when does entanglement serve as
a nonclassical resource? For our problem, we can nicely answer this question by calculating the
difference in the number of qubits needed for the same estimation error with and without
entanglement. Basically we have two limiting cases.
First, we compare the error $\bar g_F$ to the optimal case given by $\bar g_{F=1}$
and second we compare $\bar g_F$ to the error $\bar f$ for the separable case.

One easily confirms the relation
$\bar g_F(N,\vec p\,) \ge \bar g_{F=1}(N,\vec p\,)$ with equality only for
$F=1$. This states that less entanglement leads towards an larger average error, or---the other
way round---we need more qubits to obtain the same quality of our
estimation if we have less entanglement. In particular if 
we require $\bar g_{F=1}(N,\vec{p}\,)=\bar g_F(\tilde N,\vec{p}\,)$, we find 
$\tilde N=\left(\frac{3}{1-4F}\right)^2 N\ge N$ by 
comparing Eq.~\eqref{gWB} to Eq.~\eqref{g}.

However, if we want to analyze the range, in which entanglement provides a nonclassical resource,
we have to compare the mixed nonseparable case to the separable case. For this purpose let us start by
looking at the robustness of nonseparable states with respect to channel estimation. Does
any nonseparable state $\hat\rho_F$, Eq.~\eqref{werner}, provide an advantage in channel estimation? 
In other words, for which probability vectors $\vec p$ do we get
\begin{equation}
  \label{ineq1}
  \bar g_F(N=R/2,\vec p\,) \le  \bar{f}(M=R/3,\vec{p}\,)\:?
\end{equation}
Numerically, one finds a value $F_{\mathrm min}\approx 0.83$, i.\,e.\ the smallest value for which
inequality \eqref{ineq1} still holds, for $p_1=p_2=p_3\approx 0.16$. 
For $F <F_{\mathrm min}$ entangled states $\hat\rho_F$ never lead to an enhancement in estimating the channel
parameters when compared to separable states. We therefore find that the entanglement of $\hat\rho_F$
has to be quite high in order to serve as a nonclassical resource for quantum channel estimation.

Let us make this even more explicit. If we want to estimate
our Pauli channel with an average error of say 1, we require
\begin{eqnarray}\label{resreq}
   \bar{f}(M=R_{f}/3,\vec{p}\,) &\stackrel{!}{=}& 1\;, \nonumber \\
   \bar g_F(N=R_{g}/2,\vec{p}\,)  &\stackrel{!}{=}& 1
\end{eqnarray}
for separable resources $R_{f}$ and nonseparable resources $R_{g}$. By solving
these Eqs.\ for the required number of qubits $R_{f}$ and 
$R_{g}$ we are able to calculate the difference 
\begin{equation}\label{DeltaR}
  \Delta R \equiv R_{f}-R_{g}
\end{equation}
in qubit--resource requirement.
Note that the absolute value of our average estimation error, that is the right hand sides of
Eqs.~\eqref{resreq}, just gives a linear scaling factor for $\Delta R$.
As we are not interested in the absolute value of $\Delta R$, but in the sign of it, e.\,g.\
$R_{f}>R_{g}$ (estimation with separable 
states needs more resources than estimation with Werner states) or  $R_{f}<R_{g}$
(estimation with separable states needs less resources than estimation with Werner states), this is
unproblematic. We think of $\Delta R$ as being a quantity in arbitrary units and only the sign
of it matters.

As an example we consider the special Pauli channel $p_1=p_2=p_3=p$ where every error type
occurs with the same probability \footnote{This Pauli channel is equivalent to a depolarizing
channel.}.

\begin{figure}[htbp]
  \begin{center}
    \includegraphics[width=\columnwidth]{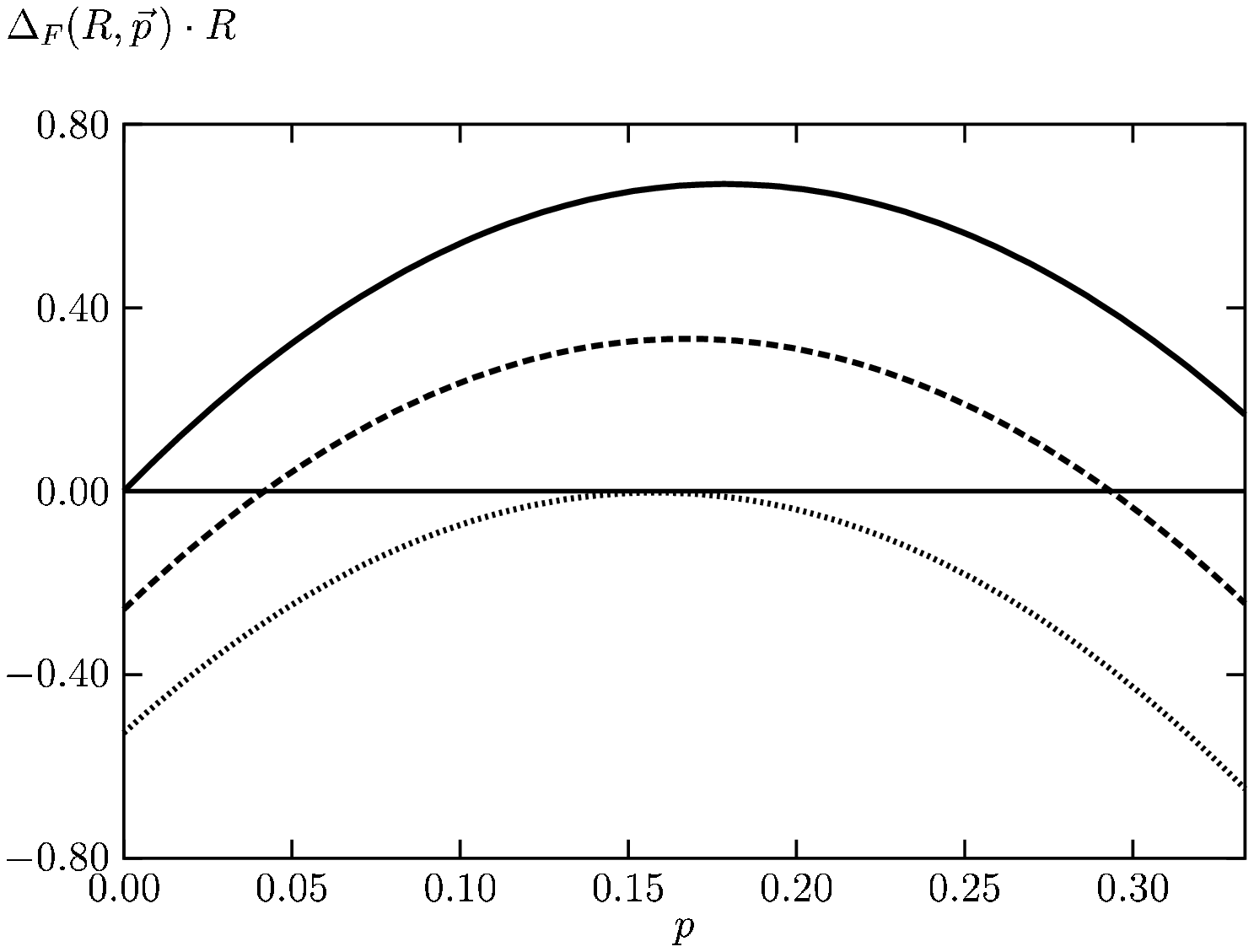}
    \vspace{0.5cm}
    \caption[Special Pauli channel]{\label{fig-gain}
     The gain $\Delta_F(R, \vec p\,)$, Eq.~\eqref{fidel-gain}, for the special Pauli
     channel $\vec p=(p,p,p)^{\mathrm T}$ plotted versus $p$. From top to bottom the graphs are
     for $F=1$ (solid line), $F=0.9$ (dashed line) and $F=F_{\mathrm min}\approx 0.83$ (dotted
     line), respectively.
     One clearly sees that the maximally entangled Bell states ($F=1$) are always superior to the
     separable case. In contrast, channel estimation with a Werner state
     $\hat\rho_{F=F_{\mathrm min}}$ leads for 
     $p=0.16$ to the same average error as the separable case, but for all other values of $p$ it is
     worse. For fidelities between these two boundaries, i.\,e.\ $F=0.9$, our nonseparable Werner
     states yield better channel estimation only for $0.04<p<0.29$.
     }
  \end{center}
\end{figure}
\begin{figure}[htbp]
  \begin{center}
    \includegraphics[width=\columnwidth]{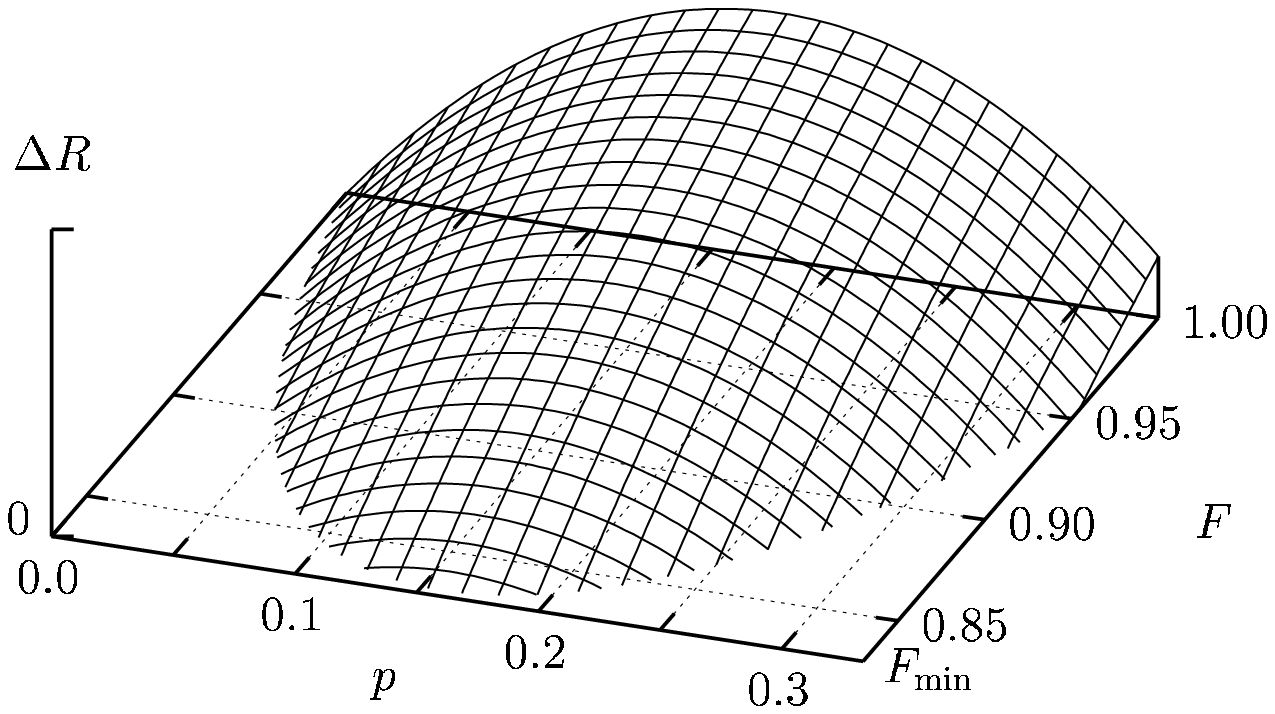}
    \vspace{0.5cm}
    \caption[Resource requirements]{\label{fig-resource}
    The difference $\Delta R$, Eq.~\eqref{DeltaR}, in the resources for the special Pauli
    channel $\vec p=(p,p,p)^{\mathrm T}$ plotted versus $p$ and $F$ in arbitrary units (see
    text). 
    There is a wide range where $\Delta R>0$. In this range we need less resources (qubits) if we
    estimate our channel with Werner states $\hat\rho_F$ as compared to an estimation with separable
    states. However, we also find a range for which $\Delta R<0$ (points not plotted) for which an
    estimation with separable states needs less resources than an estimation with Werner states.
    }
  \end{center}
\end{figure}
\begin{figure}[htbp]
  \begin{center}
    \includegraphics[scale=1]{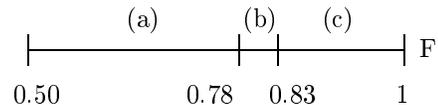}
    \vspace{0.5cm}
    \caption[Special values for the fidelity]{\label{fig-Fvalues}
      Classifications of nonseparable Werner states $\hat\rho_F$ guided by quantum channel
      estimation. In the region (a), $F>\frac{1}{2}$ the Werner state is nonseparable, and for
      $F>0.78$ in region (b) it even violates the CHSH inequality. However, only for even larger
      values $F>0.83$ in (c) $\hat\rho_F$ can be used as a nonclassical resource for
      channel estimation.
    }
  \end{center}
\end{figure}

In Fig.~\ref{fig-gain} we first show the error gain
\begin{equation}\label{fidel-gain}
  \Delta_F(R,\vec p\,) \equiv \bar f(R/3,\vec p\,)-\bar g_F(R/2,\vec p\,)
\end{equation}
which entanglement allows in contrast to separable states for different fidelities $F$.
As mentioned above, $F=1$ (solid line in Fig.~\ref{fig-gain}) always enhances the estimation, whereas for example $F=0.8$
never does so.
For $0.83 < F < 1$ it depends on the value of $\vec p$ if entanglement yields better or worse
estimation results than
the separable case. Consider for instance the case $F=0.9$ in Fig.~\ref{fig-gain} (dashed line). One
easily checks
that entangled states are only superior for $0.04 <p< 0.29$. If we know that our Pauli channel
is not parameterized by a probability vector $\vec p$ out of this domain, it is more clever to use
separable states for the estimation.
This behavior comes out most clearly when we look at the difference $\Delta R$,
Eq.~\eqref{DeltaR}, in the needed resources, shown in Fig.~\ref{fig-resource}.
Finally, in Fig.~\ref{fig-Fvalues} we shortly summarize the different important values of the Werner--state
fidelity $F$. We see that even Werner states that violate the CHSH inequality are not necessarily enhancing the
channel estimation.

\section{Estimation with Bell--diagonal states}\label{sec:Bell}

As we have seen in the preceding section that mixed nonseparable states can enhance the quality of
estimation protocols, we now extend our scheme to general Bell diagonal states and show
that channel estimation with Werner states $\hat\rho_F$, Eq.~\eqref{werner}, leads to the smallest average error, when nothing about
the error probabilities of the Pauli channel is known.

We consider the Bell diagonal state
\begin{eqnarray}\label{rB}
  \hat\rho_{\mathrm B} &=&\alpha_1 \bpsimi + \alpha_2 \bpsipl \nonumber \\
                       &+&\alpha_3 \bphimi +\alpha_4 \bphipl\;,
\end{eqnarray}
with normalization condition $\sum_{i=1}^4 \alpha_i=1$, $\alpha_i\ge 0$. Without loss of generality, we
assume $\alpha_1 \ge\alpha_{2}, \alpha_{3}, \alpha_{4}$. For $\alpha_1>\frac{1}{2}$ this state is
nonseparable \cite{vidal} and has the same fidelity
$F=\langle\psi^-|\hat\rho_{\mathrm B}|\psi^-\rangle=\alpha_1$ as our previously considered Werner
state $\hat\rho_F$. Therefore, we keep 
$\alpha_1$ fixed in order to compare the average error in the Werner case $\bar g_F(N,\vec{p}\,)$, Eq.~\eqref{gWB},
with the average error in the Bell diagonal case $\bar g_{\mathrm B}(N,\vec{p}\,)$.

As in the previous case, Alice's qubit is sent through the channel while Bob simply keeps his
qubit. Performing a joint Bell measurement at Bob's site enables us to estimate the channel probabilities $\vec p$ and
calculate the average error denoted by $\bar g_{\mathrm B}(N,\vec{p}\,)$.

Due to the fact that the general expression for
$\bar g_{\mathrm B}(N,\vec{p}\,)$ is lenghty and rather complex, we do not present the explicit expression
here. It is more interesting to look at the mean error $\langle \bar g_{\mathrm B}(N)\rangle_{\vec p}\,$
averaged over all possible Pauli channels. For this mean error we find
\begin{eqnarray}\label{gInt}
  \langle \bar g_{\mathrm B} (N)\rangle_{\vec p} 
             &=&\iiint \limits_{0\le p_1+p_2+p_3\le 1} \bar g_{\mathrm B}(N,\vec{p}\,) \;{\mathrm d} p_1\,{\mathrm d} p_2\,{\mathrm d} p_3 \nonumber \\
             &=&\frac{1}{32 N}\left(\frac{1}{(1-2\alpha_1-2\alpha_2)^2} + \frac{1}{(1-2\alpha_1-2\alpha_3)^2} \right.\nonumber \\
             &&\qquad\quad \left. +\frac{1}{(1-2\alpha_2-2\alpha_3)^2} - \frac{3}{5}\right)\;.
\end{eqnarray}
For fixed $\alpha_1$ the mean error $\langle \bar g_{\mathrm B} (N)\rangle_{\vec p}$ has a global minimum at
\begin{equation}
 \alpha_2=\alpha_3=\alpha_4=\frac{1-\alpha_1}{3}\;.
\end{equation}
This means that our Bell diagonal state, Eq.~\eqref{rB}, leads to the minimal error if it is in a Werner state.
Note that a general Bell diagonal state can always be transformed to a Werner state by randomly applying bilateral
rotations \cite{bennett}.

If we do not know anything about our Pauli channel and only have a specific amount
of entanglement available, then it is best to use a Werner
state for estimating $\vec p$. However, as shown above, the fidelity $F=\alpha_1$ needs to be quite high in order to beat the
estimation scheme with separable states.


\section{Conclusions}\label{sec:Concl}
Entanglement serves as a superior resource for Pauli channel estimation \cite{fischer}. This
nonclassical resource enables us to estimate the parameters of a Pauli channel with an lower error
than in the classical way or, in other words, we need less entangled test qubits than separable test
qubits to arrive at the same estimation error. However for the discussed application, we have seen
that entanglement is not very robust. We need a high amount of entanglement to have a chance to
profit from this nonseparable resource. And 
even if the fidelity of our Werner state is high enough, it depends on the actual channel parameters
if we can benefit from mixed--state entanglement.
We have also shown that Werner states are optimal for estimation in the sense that they yield the
lowest average estimation error when compared to a general Bell--diagonal state.

\section*{Acknowledgments}

We thank A.~Delgado and M.A.~Cirone for fruitful discussions. This work was supported by the DFG
program ``Quanten--Informationsverarbeitung'', by the European Science Foundation QIT program, by
the IST program ``QUBITS'' and the IHP network ``QUEST'' of the European Commission.


{\appendix
\section{The \lowercase{$d$}--dimensional case}\label{sec:d-dim-werner}

The Werner state estimation scheme for the Pauli channel can easily be extended to $d$ dimensions.
This extension of the Pauli channel to higher dimensional Hilbert spaces has recently been studied
in the context of quantum error correction, quantum cloning machines and entanglement \cite{genpauli}.

The channel errors in $d$ dimensions can be described by the unitary transformations
\begin{equation}
  \hat U_{m,n}=\sum_{k=0}^{d-1}{\mathrm e}^{2\pi i (kn/d)} |k+m\rangle\langle k|\;,
\end{equation}
with $\hat U_{0,0}=\hat 1$ being the identity in a Hilbert space spanned by the orthonormal basis
states $|0\rangle$, $|1\rangle$, \ldots, $|d-1\rangle$.
The generalized Pauli channel then reads
\begin{equation}\label{app:rho}
  C(\hat \rho)=\sum_{m,n=0}^{d-1} \pmn\,\hat U_{m,n}\,\hat\rho\,\hat U_{m,n}^\dagger\;
\end{equation}
with error probabilities $\pmn$ ($p_{0,0}$ is the probability for no error),
$\sum_{m,n=0}^{d-1}\pmn=1$. We further define the maximally entangled states
\begin{equation}
  \psimn = \frac{1}{\sqrt{d}} \sum_{j=0}^{d-1} {\mathrm e}^{2\pi i (jn/d)} |j\rangle|j+m\rangle\;.
\end{equation}
and the $d$--dimensional Werner state can then be expressed in the form
\begin{equation}
  \hat \rho_\lambda=(1-\lambda)\frac{1}{d^2}\hat 1+\lambda |\psi_{0,0}\rangle\langle\psi_{0,0}|\;.
\end{equation}
Again the fidelity
\begin{equation}\label{d-F}
  F=\langle\psi_{0,0}| \hat \rho_\lambda |\psi_{0,0}\rangle = \lambda + \frac{1-\lambda}{d^2}\;.
\end{equation}
is defined as the overlap of $\hat\rho_\lambda$ with respect to $|\psi_{0,0}\rangle\langle\psi_{0,0}|$.
As in the two--dimensional case, Alice sends her qubits through the channel while Bob simply keeps
his qubits. A generalized Bell measurement on the output state $C(\hat\rho_\lambda)$,
Eq.~\eqref{app:rho}, now 
corresponds to the set of orthonormal states $\psimn$. Hence one finds $i_{m,n}$ times the
state $\psimn$ ($\sum_{m,n=0}^{d} i_{m,n}=N$). From these measured frequencies the estimated channel
parameters $\pmn^{\mathrm est}$ can now be calculated via
\begin{equation}
  \pmn^{\mathrm est}=\frac{i_{m,n}-\frac{1-\lambda}{d^2}}{\lambda}\:.
\end{equation}
Consequently the average error takes the form
\begin{eqnarray}
 \bar g_\lambda(N,\{\pmn\}))
         &=& \frac{1}{N} \frac{1}{\lambda^2} {\sum_{m,n}}'
         \Bigl\{ \pmn(1-\pmn) \nonumber \\
         && \quad -\, (1-\lambda)\left(\pmn-\frac{1}{d^2}\right)(1-2 \pmn) \nonumber \\
         && \quad -\, (1-\lambda)^2\left(\pmn-\frac{1}{d^2}\right)^2 \Bigr\}\;,
\end{eqnarray}
where ${\sum_{m,n}}'$ denotes summation over $m$, $n$ from 0 to $d-1$ omitting the pair $m=n=0$.

If we set $d=2$ and insert Eq.~\eqref{d-F} we of course obtain the average error, Eq.~\eqref{gWp},
from Sec.~\ref{sec:EstSchem}. It is interesting to note that the error probabilities
$\pmn$ do not appear in combinations like $\pmn \cdot p_{m',n'}$ where $m' \ne m$ and $n' \ne
n$. This is due to the fact that we only send a qubit once through the channel. Each time our channel
superoperator $C(\hat \rho_\lambda)$ just acts as one error operator and therefore each qubits gets one error probability `attached'.

} 




\begin{thebibliography}{00}
\bibitem{qip}
For recent books on the subject, see
H.-K. Lo, T. Spiller and S. Popescu (eds.), {\it Introduction to Quantum Computation and Information}, World Scientific Publishing, Singapore, 1998;
J.~Gruska, {\it Quantum Computing}, McGraw Hill, London, 1999;
D.~Bouwmeester, A.~Ekert and A.~Zeilinger (eds.), {\it The Physics of Quantum Information}, Springer, Berlin, 2000;
M.\,A.~Nielsen and I.\,L.~Chuang, {\it Quantum Computation and Quantum Information},
Cambridge University Press, Cambridge, 2000;
G.~Alber, T.~Beth, M.~Horodecki, P.~Horodecki, R.~Horodecki, M.~R\"otteler, H.~Weinfurter, R.~Werner,
A.~Zeilinger, {\it Quantum Information: An Introduction to Basic Theoretical Concepts and Experiments},
Springer, Berlin, 2001.
\bibitem{fundres}
A.~Einstein, B.~Podolsky, and N.~Rosen, Phys.~Rev.\ {\bf 47}, 777 (1935);
J.\,S.~Bell, Physics {\bf 1}, 195 (1964);
D.\,M.~Greenberger, M.\,A.~Horne, and A.~Zeilinger, in {\it Bell's Theorem. Quantum Theory, and
  Conceptions of the Universe}, edited by M.~Kafatos (Kluwer, Dordrecht, 1989), p.~69.
\bibitem{teleport} C.\,H.~Bennett, G.~Brassard, C.~Cr\'epeau, R.~Jozsa, A.~Peres, and W.\,K.~Wootters, Phys. Rev. Lett. {\bf 70}, 1895 (1993);
D.~Bouwmeester, J.-W.~Pan, K.~Mattle, M.~Eibl, H.~Weinfurter, and A.~Zeilinger, Nature {\bf 390}, 575 (1997);
D.~Boschi, S.~Branca, F.~De~Martini, L.~Hardy, and S.~Popescu, Phys.~Rev.~Lett. {\bf 80}, 1121 (1998);
A.~Furusawa, J.\,L.~S{\o}rensen, S.\,L.~Braunstein, C.\,A.~Fuchs, H.\,J.~Kimble, and E.\,S.~Polzik, Science {\bf 282}, 706 (1998);
M.\,A.~Nielsen, E.~Knill, and R.~Laflamme, Nature {\bf 396}, 52 (1998).
\bibitem{qucrypt}
C.\,H.~Bennett and G.~Brassard, in {\it Proceedings of IEEE International Conference on Computers, Systems and Signal Processing}, IEEE, New York (1984);
A.~Muller, J.~Breguet, and N.~Gisin, Europhys.~Lett.~ {\bf 23}, 383 (1993);
A.~Ekert, Phys.~Rev.~Lett. {\bf 67}, 661 (1991).
\bibitem{horodecki}
C.\,H.~Bennett, G.~Brassard, S.~Popescu, B.~Schumacher, J.\,A.~Smolin, and W.\,K. Wootters,
Phys.~Rev.~Lett. {\bf 76}, 722 (1996);
S.~Popescu and D.~Rohrlich, Phys.~Rev.~A {\bf 56}, R3319 (1997);
V.~Vedral, M.\,B.~Plenio, M.\,A.~Rippin, and P.\,L.~Knight, Phys.~Rev.~Lett. {\bf 78}, 2275 (1997);
V.~Vedral and M.\,B.~Plenio, Phys.~Rev.~A {\bf 57}, 1619 (1998).
\bibitem{horodecki-rev}
For a recent review on entanglement measures, see M.~Horodecki, Quantum Information and Computation {\bf 1}, 3 (2001).
\bibitem{fischer}
D.\,G.~Fischer, H.~Mack, M.\,A.~Cirone, and M.~Freyberger, Phys.~Rev.~A {\bf 64}, 022309 (2001).
\bibitem{vidal}
G.~Vidal and R.\,F.~Werner, e--print quant-ph/0102117 (2001).
\bibitem{werner}
R.\,F.~Werner, Phys.~Rev.~A {\bf 40}, 4277 (1989).
\bibitem{bennett}
C.\,H.~Bennett, D.\,P.~DiVincenzo, J.\,A.~Smolin, and W.\,K.~Wootters, Phys.~Rev.~A {\bf 54}, 3824 (1996).
\bibitem{chsh}
J.\,F.~Clauser, M.\,A.~Horne, A.~Shimony, and R.\,A.~Holt, Phys.~Rev.~Lett. {\bf 23}, 880 (1969).
\bibitem{peres}
A.~Peres, Phys.~Rev.~A {\bf 54}, 2685 (1996).
\bibitem{genpauli}
E.~Knill, e--prints quant-ph/9608048 and quant-ph/9608049 (1996);
D.~Gottesman, Chaos, Solitons, and Fractals {\bf 10}, 1749 (1999);
N.\,J.~Cerf, J.~Mod.~Opt. {\bf 47}, 187 (2000).
\end{thebibliography}
\end{document}